\documentclass[12pt]{iopart}
\usepackage{iopams}  
\usepackage{stix}
\usepackage[utf8]{inputenc}
\usepackage{graphicx}
\begin{document}

\title{Emergence of a spectral gap in a class of random matrices associated with split graphs}

\author{Kevin E. Bassler$^{1,2,3,4}$ and R.K.P. Zia$^{4,5,6}$}
\address{$^{1}$ Department of Physics, University of Houston, Houston, TX 77204, USA.}
\address{$^{2}$ Department of Mathematics, University of Houston, Houston, TX 77204, USA.}
\address{$^{3}$ Texas Center for Superconductivity, University of Houston, Houston, TX 77204, USA.}
\address{$^{4}$ Max-Planck-Institut f\"{u}r Physik komplexer Systeme, N\"{o}thnitzer
Str. 38, Dresden D-01187, Germany} 
\address{$^{5}$ Center for Soft Matter and Biological Physics, 
Physics Department, Virginia Polytechnic Institute and State University, Blacksburg, VA 24061, USA} 
\address{$^{6}$ Department of Physics and Astronomy, Iowa State University,
Ames, IA 50011, USA}
\date{2017}

\ead{\mailto{bassler@uh.edu},\mailto{rkpzia@vt.edu}}
\begin{abstract}
Motivated by the intriguing behavior displayed in a dynamic network that models a population of extreme introverts and extroverts (XIE), we consider the spectral properties of ensembles of random split graph adjacency matrices. 
We discover that, in general, a gap emerges in the bulk spectrum between $-1$ and $0$ that contains a single eigenvalue. An analytic expression for the bulk distribution is derived and verified with numerical analysis. We also examine their relation to chiral ensembles, which are associated with bipartite graphs.
\end{abstract}

\maketitle

\section{Introduction}
The properties of ensembles of random matrices, particularly the spectra of their eigenvalues, are important for many applications and have been the basis of foundational studies in fields ranging from physics~\cite{PorterBook} to statistics~\cite{W1928} to ecology~\cite{May1972,May1973}.
They have also been of broad interest in both theoretical physics and pure mathematics~\cite{FSB2003,ForBook}.
Among the most notable results in random matrix theory is the
Wigner semi-circle law~\cite{Wigner1955,Wigner1958}. It applies to classes of matrices including real symmetric  matrices  and complex hermitian  matrices with  i.i.d. random diagonal elements and i.i.d. random  off-diagonal elements, which can be different distributions, so long as the moments of the off-diagonal distribution, up through the fourth one, are finite~\cite{a1971}. For such matrix ensembles, Wigner's law says that in the large matrix limit, the  eigenvalue probability distribution function (epdf) is a semi-circle. Similar  results apply to other classes of random  matrices.  The singular values of asymptotically large rectangular matrices having i.i.d. random elements with finite variance are distributed according to the  Marchenko-Pastur law~\cite{MP1967}.  Also, the eigenvalues of chiral random matrices come in plus/minus pairs, but have a semi-circle pdf in the large matrix limit~\cite{ForBook}.
Low-rank perturbations of random matrices, however, can cause epdfs to deviate from these semi-circle
type laws in interesting ways when the matrices are finite~\cite{ForBook,BGN2011,OW2017}.

In network science, eigenvalues of ensembles of random matrices are used to
characterize the structure and properties of classes of networks, or graphs~%
\cite{cheung1994,BJ2007,mieghem2011}. 
Spectra can reveal structural properties including modularity~\cite{nn2012,p2013} and existence of motifs~\cite{MSA2008,NGB2015,DK2017}
that affect dynamical properties including synchronizability~\cite{ADKYZ2008} and redundancy~\cite{MS2009}.
Adjacency and Laplacian matrices both provide complete
descriptions of the structure of networks. Studies of ensembles of random ensembles
adjacency matrices have shown that the largest eigenvalue typically separates from the bulk distribution~\cite{clv2003} due to
the non-zero mean value of the elements~\cite{L1964,JKT1968,BFF2009} and that the bulk of the eigenvalue distribution of Erdős-Rényi networks~\cite{er1959} follows the semi-circle law
in the limit of infinite mean degree~\cite{FK1981,RCKT2008},
but that it can have a different, ``triangle-like'' distribution with power-law tails when the degree of the nodes are power-law distributed~\cite{clv2003,fdbv2001}.

In this context, our recent interest in a class of non-equilibrium dynamical models
of social networks motivated us to study the ensembles of their associated
adjacency matrices. In these models~\cite{ZLJS2011, ZLS2012,
LSZ2012,LJSZ2013, LSZ2014}, a node (an individual in a social setting) may
add or cut edges (contacts with other individuals) according to some
\textquotedblleft preference.\textquotedblright\ Such propensities may be
internal, such as introverts preferring few contacts and extroverts, many.
There may also be external circumstances which affect how many contacts an
individual might \textquotedblleft prefer,\textquotedblright\ e.g., during a
raging epidemic. Assuming time independent probabilities for individuals to
cut/add links, the system will come into steady state, leading to a
time-independent ensemble of adjacency matrices, $\mathbb{A}$. Typically,
the dynamics governing the evolution of such systems do not obey detailed
balance~\cite{LJSZ2013} and so, the stationary distribution, 
$\mathcal{P}^{\ast }\left( \mathbb{A}\right) $ is not known, 
so that Monte Carlo simulations provide
the only way to proceed. Remarkably, there is a special limit where detailed
balance holds and an explicit $\mathcal{P}^{\ast }\left( \mathbb{A}\right) $
is found~\cite{LSZ2012}. Known as the XIE model, this limit consists of a population of
extreme introverts (I) and extroverts (E), in which an I, when chosen, cuts
a randomly chosen link while a chosen E adds a link to a random individual
not already connected to it. As a result, the $\mathbb{A}$'s reduce to
blocks, corresponding to no I-I links, fully connected E-E group, and a
dynamic set of I-E cross-links.

In the context of graph theory, such networks are associated with split
graphs \cite{FH1977,TC1979}, with many interesting properties. Thus, our focus here
can also be phrased as \textquotedblleft properties of the spectra of
split graph adjacency matrices.\textquotedblright\ To be
specific, consider the ensemble of $2N\times 2N$ matrices $\mathbb{A}$ of
the form 
\begin{equation}
\mathbb{A}=\left( 
\begin{array}{cc}
\mathbb{0} & \mathbb{X} \\ 
\mathbb{X}^{T} & \mathbb{M}%
\end{array}%
\right)   \label{A}
\end{equation}%
where $\mathbb{X}$ is an $N\times N$ square matrix with i.i.d. random
Boolean variables ($0$ or $1$) as elements and $\mathbb{0}$ is a matrix with
only $0$ elements. Here, 
\begin{equation}
\mathbb{M=-I}+\left\vert u\rangle \langle u\right\vert   \label{M-def}
\end{equation}%
where $\left\vert u\right\rangle $ is the ($N\times 1$) vector with unity in
every element\footnote{%
Note that, for any matrix $\mathbb{A}$, $\left\langle u\right\vert \mathbb{A}%
\left\vert u\right\rangle $ is the sum over all its elements.} and $%
\left\langle u\right\vert =\left\vert u\right\rangle ^{T}$. Matrices of this
form appear naturally in \textquotedblleft critical\textquotedblright\ XIE
networks \cite{BLSZ2015,BDZ2015}, i.e., ones with with equal numbers ($N$)
of introverts and extroverts. Thus, we will refer to matrices of the
above form \textit{XIE matrices. }To reiterate, the $0$ block diagonal
matrix corresponds to the adjacency matrix of the I's, $\mathbb{M}$ the
adjacency matrix of the E's, and $\mathbb{X}$ the incidence matrix of the
I-E pairs. Now, there are non-trivial correlations among the I-E
cross-links, so that the elements of $\mathbb{X}$ are not i.i.d. 
The differences between the epdf of such adjacency matrices and that 
of random split graphs should reflect the correlations
between the elements in $\mathbb{X}$. In this paper, as a first step, we restrict consideration to random split graphs. 

\begin{figure}
\centering
\includegraphics[width=\textwidth]{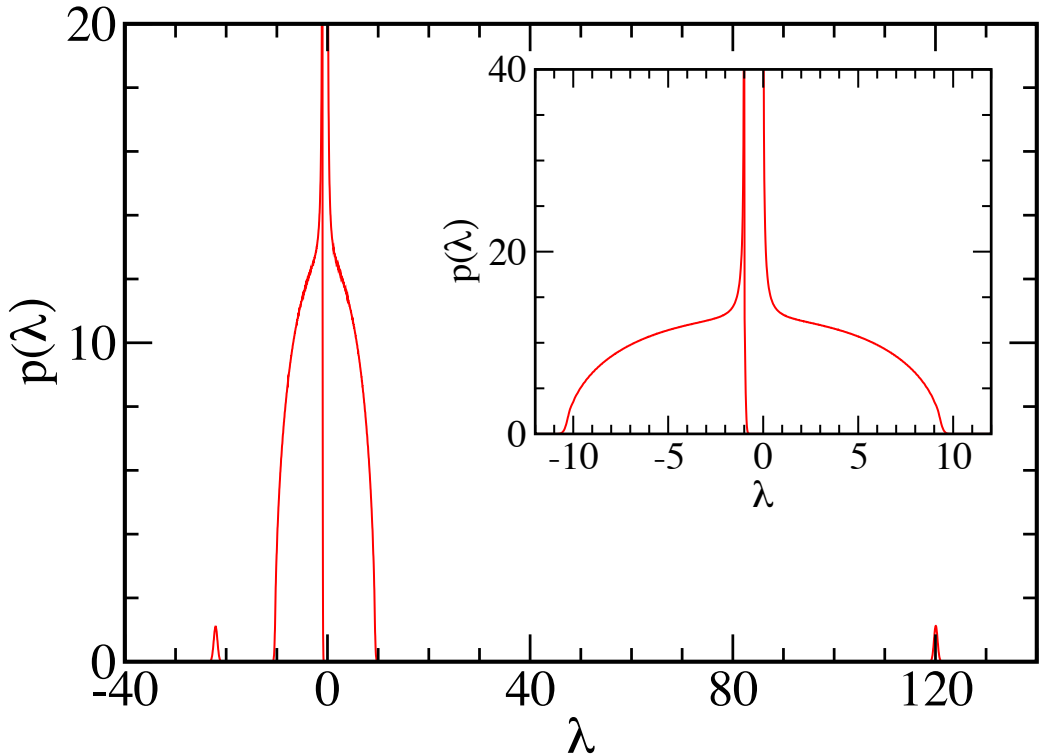}
\caption{Eigenvalue probability distribution function of $2N \times 2N$ adjacency matrices of random split graphs with $N=100$
and I-E connection probability of 0.5. 
Note the two separated single eigenvalues, near $-22$ and $120$. Inset shows an enlarged view of the bulk distribution, which resembles the steeple of Stykkish\'{o}lmskirkja. 
A gap between $-1$ and $0$ is clearly visible. Further, there is a single eigenvalue \textquotedblleft tightly bound\textquotedblright\ to $-1$ (in this case), a fact evidenced by the almost vertical line at the left edge of the gap.
}
\label{gap-fig1}
\end{figure}

Closely related to split graphs are bipartite graphs, corresponding to a
network with only I-E links. The associated adjacency matrices are known as 
\textit{chiral} matrices\footnote{%
In most of the literature, chiral matrices are of the form $\left( 
\begin{array}{cc}
\mathbb{0} & \mathbb{X} \\ 
-\mathbb{X}^{T} & \mathbb{0}%
\end{array}%
\right) $. The ones here differ only by a sign, i.e., $\left( 
\begin{array}{cc}
\mathbb{0} & \mathbb{X} \\ 
\mathbb{X}^{T} & \mathbb{0}%
\end{array}%
\right) $. The two are clearly intimately related, with properties that can
be mapped from one to the other. We follow the terminology of \cite{ForBook} where the latter are referred to as chiral matrices.}. The
principal finding reported here is the emergence of a gap in the epdf of
such matrices when \textquotedblleft $\mathbb{M}$ is
added,\textquotedblright\ i.e., when a bipartite graph is connected to form
a split graph (or when one of the two sets of independent nodes are linked
to form a clique). In particular, for chiral matrices with random Boolean
variables ($0$ or $1$) of equal probability as elements, the epdf is symmetric around $0$, with the bulk obeying the simple semi-circle
law, apart from isolated ev's far outside \cite{BFF2009,BFF2010}. By contrast, Fig.~\ref{gap-fig1}
shows the striking epdf associated with the latter, reminiscent of  a (upside down) Viking ship or the steeple of the Stykkish\'{o}lmskirkja. 
Many
different features are evident: asymmetry, serious distortions from
semi-circle, peaks at -1 and 0, and a gap between them. It appears
as if the negative part of the distorted semi-circle were shifted to more
negative values, while the isolated eigenvalues on edge of the bulk appears
to be shifted to more positive values! Additionally, careful examination
shows a single eigenvalue lying within the gap (close to $-1$ here). 
The two eigenvalues separated from the bulk appear as isolated peaks (near $-22$
and $120$ here). By contrast, the eigenvalue in the gap appears
\textquotedblleft tightly bound\textquotedblright\ to the left edge, so that
the pdf descends rapidly as $\lambda $ increases from $-1$. For more generic 
$\mathbb{X}$'s (e.g., different means), this eigenvalue can be detached from
both ends and isolated. All of these remarkable features will be examined
and explained.
Random split graphs are a type of stochastic block model~\cite{HLH1983,FMW1985,FW1992} for which 
methods of spectral analysis are well developed~\cite{p2013}. However, those methods require
a finite variance for the elements of the blocks. The elements of the I-I and E-E blocks
of split graph adjacency matrices have zero variance and, so, require other methods of analysis.  

The remainder of this paper is organized as follows. We begin by establishing
the connection between the adjacency matrices of bipartite and
a split graphs, and their relation to chiral matrices. 
Through these connections, many of the features shown in Fig.~\ref{gap-fig1} can be understood and,
in the case where $\mathbb{X}$ consists of i.i.d. Gaussian variables of zero
mean, the epdf can be computed analytically (in the large $N$ limit). We then
present simulation results for various
generalizations: shifted mean Gaussians, Boolean ($0,1$) distributions, and
split graph adjacency matrices. We end with a brief summary and outlook for
further studies. The Appendix is devoted to some details associated with
special cases.

\begin{figure}
\centering
\includegraphics[width=\textwidth]{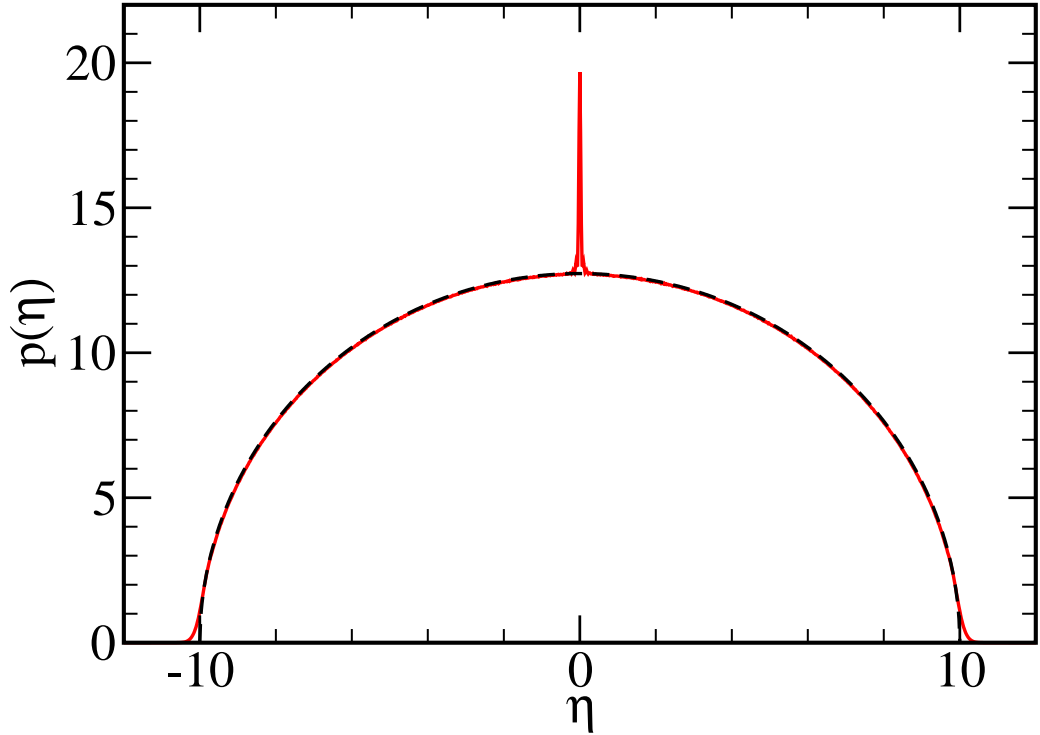}
\caption{Eigenvalue probability distribution function of $2N \times 2N$ random chiral matrices with $N=100$
and Gaussian distributed elements with zero mean and variance of 0.25.  
Black dashed line is the corresponding Wigner semi-circle prediction.
}
\label{gap-fig2}
\end{figure}

\section{Results}

This section is devoted to establishing the connection between the spectra of chiral
and XIE matrices, i.e., matrices of the form (4 blocks of $N\times N$
matrices) 
\begin{equation}
\mathbb{A}_{B}=\left( 
\begin{array}{cc}
\mathbb{0} & \mathbb{X} \\ 
\mathbb{X}^{T} & \mathbb{0}%
\end{array}%
\right) \qquad
\mbox{and}
\qquad \mathbb{A}_{S}=\left( 
\begin{array}{cc}
\mathbb{0} & \mathbb{X} \\ 
\mathbb{X}^{T} & \mathbb{M}%
\end{array}%
\right).  \label{AB+AS}
\end{equation}
When $\mathbb{X}$ has Boolean elements, these forms correspond to adjacency matrices of, respectively, bipartite graphs (B) and split graphs (S).  However, in what follows, we also consider matrices with $\mathbb{X}$s that have Gaussian distributed elements. We begin with finding the connection between these matrices for a specific $\mathbb{X}$. Then, we consider ensembles of them with random i.i.d. elements, first examining Gaussians distributions with zero mean and then exploring ones with positive mean, specifically, with Boolean elements.  

\subsection{Considerations for a specific $\mathbb{X}$\label{specX}}

First, consider a \textit{particular} $\mathbb{X}$ (as opposed to an
ensemble of them), and denote the eigenvectors of $\mathbb{A}_{B}$ by the
doublet of $N\times 1$ vectors along with eigenvalues $\eta _{\alpha }$
\begin{equation}
\mathbb{A}_{B}\left( 
\begin{array}{c}
\left\vert \tilde{w}_{\alpha }\right\rangle \\ 
\left\vert w_{\alpha }\right\rangle%
\end{array}%
\right) =\eta _{\alpha }\left( 
\begin{array}{c}
\left\vert \tilde{w}_{\alpha }\right\rangle \\ 
\left\vert w_{\alpha }\right\rangle%
\end{array}%
\right)  \label{doublet}
\end{equation}%
Thus, 
\begin{equation}
\mathbb{X}\left\vert w_{\alpha }\right\rangle =\eta _{\alpha }\left\vert 
\tilde{w}_{\alpha }\right\rangle ;~~\mathbb{X}^{T}\left\vert \tilde{w}%
_{\alpha }\right\rangle =\eta _{\alpha }\left\vert w_{\alpha }\right\rangle
\label{X eqn}
\end{equation}%
so that%
\begin{equation}
\mathbb{W}\left\vert w_{\alpha }\right\rangle \equiv \mathbb{X}^{T}\mathbb{X}%
\left\vert w_{\alpha }\right\rangle =\mu _{\alpha }\left\vert w_{\alpha
}\right\rangle  \label{W ev}
\end{equation}%
with%
\begin{equation*}
\mu _{\alpha }\equiv \eta _{\alpha }^{2};~~\alpha =1,...,N
\end{equation*}%
Note that $\mathbb{W}$ is referred to as a Wishart matrix \cite{ForBook} 
\footnote{%
The collection of $\mathbb{W}$'s with random $\mathbb{X}$'s is known as the
Laguerre ensemble.}. For later convenience, we will assume the $w$'s are
normalized, so that any function of the matrix $\mathbb{W}$ can be written as%
\begin{equation}
f\left( \mathbb{W}\right) =\sum_{\alpha }f\left( \mu _{\alpha }\right)
\left\vert w_{\alpha }\right\rangle \langle w_{\alpha }|  \label{f(L)}
\end{equation}%
We will also focus on the generic case where there are no degeneracies and
strictly positive $\mu $'s: $0<\mu _{1}\leq \mu _{2}\leq ...\leq \mu _{N}$.
Exceptions to these will be deferred to the Appendix. From here, it is easy
to see that the $2N$ $\eta $'s come in $N$ plus/minus pairs, which we label by $\eta _{\alpha \pm
}=\pm \sqrt{\mu _{\alpha }}$. Associated with these are eigenvectors (of $%
\mathbb{A}_{B}$) in the form of a doublet: $\left(\mathbb{X}\left\vert w_{\alpha
}\right\rangle \left/ \eta _{\alpha \pm }\right. ,\left\vert w_{\alpha
}\right\rangle \right)$.

Proceeding, we consider the eigenvectors and eigenvalues of $\mathbb{A}_{S}:$
\begin{equation}
\mathbb{A}_{S}\left( 
\begin{array}{c}
\left\vert \tilde{v}\right\rangle \\ 
\left\vert v\right\rangle%
\end{array}%
\right) =\lambda \left( 
\begin{array}{c}
\left\vert \tilde{v}\right\rangle \\ 
\left\vert v\right\rangle%
\end{array}%
\right)  \label{evEqn-As}
\end{equation}%
Consider the case $\lambda \neq 0$ first so that we may use $\left\vert 
\tilde{v}\right\rangle =\mathbb{X}\left\vert v\right\rangle /\lambda $ to
eliminate $\left\vert \tilde{v}\right\rangle $ as before and arrive at%
\begin{equation}
\mathbb{W}\left\vert v\right\rangle /\lambda +\mathbb{M}\left\vert
v\right\rangle =\lambda \left\vert v\right\rangle
\end{equation}%
Inserting the explicit form (\ref{M-def}), we find%
\begin{equation}
\mathbb{W}\left\vert v\right\rangle =\left( \lambda ^{2}+\lambda \right)
\left\vert v\right\rangle -\lambda c\left\vert u\right\rangle
\label{EigenEqn}
\end{equation}%
where $c\equiv \left\langle u|v\right\rangle $ is just the sum of the
elements of $\left\vert v\right\rangle $. The solution to (\ref{EigenEqn})
is clear%
\begin{equation}
\left\vert v\right\rangle =\lambda c\left[ \lambda ^{2}+\lambda -\mathbb{W}%
\right] ^{-1}\left\vert u\right\rangle  \label{v}
\end{equation}%
Projecting onto $\left\vert u\right\rangle $, we have $c=\lambda c\langle u|%
\left[ \lambda ^{2}+\lambda -\mathbb{W}\right] ^{-1}\left\vert
u\right\rangle $. Assuming\footnote{%
The special cases where $\lambda $ or $c$ vanish are also studied in the
Appendix.} $c\neq 0$ and exploiting (\ref{f(L)}), an explicit form for the
secular equation emerges:%
\begin{equation}
1=\lambda \sum_{\alpha =1}^{N}\frac{c_{\alpha }^{2}}{\lambda ^{2}+\lambda
-\mu _{\alpha }}  \label{evsEqn}
\end{equation}%
where 
\begin{equation*}
c_{\alpha }\equiv \left\langle u|w_{\alpha }\right\rangle
\end{equation*}%
The right hand side can be written as a sum, 
\begin{equation}
1=\sum_{\alpha }\frac{c_{\alpha }^{2}}{\rho _{\alpha +}-\rho _{\alpha -}}%
\left[ \frac{\rho _{\alpha +}}{\lambda -\rho _{\alpha +}}+\frac{\left( -\rho
_{\alpha -}\right) }{\lambda -\rho _{\alpha -}}\right]  \label{master}
\end{equation}%
over $2N$ simple poles, at%
\begin{equation}
\rho _{\alpha \pm }=\frac{1}{2}\left\{ -1\pm \sqrt{1+4\mu _{\alpha }}\right\}
\label{lambdas}
\end{equation}%
Since $\mu >0$, these poles are located \textit{outside} the interval $\left[
-1,0\right] $, while all residues are \textit{positive}. Thus, as $\lambda $
is varied from $-\infty $ to $+\infty $, the right hand side begins at $0$,
falls to $-\infty $ at $\rho _{N-}=\left( -1-\sqrt{1+4\mu _{N}}\right) /2$,
runs from $+\infty $ to $-\infty $ between each successive $\rho _{\alpha \pm
}$ until past $\rho _{N+}=\left( -1+\sqrt{1+4\mu _{N}}\right) /2$, and
finally falls from $+\infty $ to $0$. The consequence is that there is
precisely one solution to (\ref{master}) between successive pairs of $\rho$, 
plus one more beyond $\rho _{N+}$. As $\alpha \in \left[ 1,N\right] $, 
there are precisely $2N$ solutions, which we label by $\lambda
_{\alpha \pm }$. Such interlacing between $\lambda $'s and $\rho $'s%
\footnote{%
Some $\lambda $'s may be equal to $\rho $'s. See Appendix 1 for details.}%
\begin{equation*}
\rho _{N-}<\lambda _{N-}<...<\rho _{1-}<\lambda _{1-}<\rho _{1+}<\lambda
_{1+}<\rho _{2+1}<...<\rho _{N+}<\lambda _{N+}
\end{equation*}
is a familiar occurrence in random matrix theory (e.g., in \cite{BFF2009}).
Of course, it is not easy to find the precise location of each $\lambda
_{\alpha }$, as details such as $\left\{ c_{\alpha }\right\} $ will
determine whether a $\lambda _{\alpha }$ is closer to $\rho _{\alpha }$ or $%
\rho _{\alpha +1}$. As will be shown below, we can nevertheless draw
meaningful conclusions for many aspects of he epdfs of random matrices.
Finally, the eigenvectors associated with $\lambda _{\alpha \pm }$ are the
doublets: $\left(\mathbb{X}\left\vert v_{\alpha \pm }\right\rangle /\lambda
_{\alpha \pm },\left\vert v_{\alpha \pm }\right\rangle \right)$, where $\left\vert
v_{\alpha \pm }\right\rangle $ is given by (\ref{v}) with $\lambda =\lambda
_{\alpha \pm }$.

\begin{figure}
\centering
\includegraphics[width=0.9\textwidth]{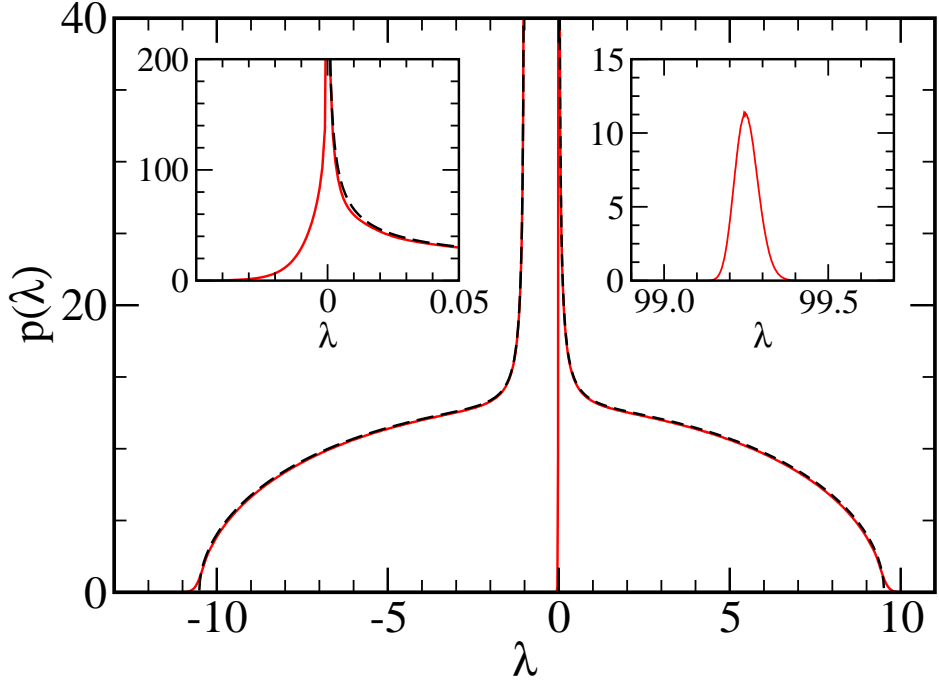}
\caption{Eigenvalue probability distribution function of $2N \times 2N$ XIE matrices with $N=100$
and I-E connection probability of 0.5
and Gaussian distributed I-E elements with zero mean and variance of 0.25.  
Black  dashed line is our analytic prediction for the bulk distribution.
Left inset shows an enlarged view of the gap region near $\lambda=0$. 
Note that the gap contains a single eigenvalue, which happens to be 
\textquotedblleft tightly bound\textquotedblright\ to $0$ in this case. 
The presence of this eigenvalue and its location is emphasized by the almost vertical line near 
the right edge of the gap in the main plot.
Right inset shows the  separated eigenvalue near $\lambda=99.25$
}
\label{gap-fig3}
\end{figure}

\subsection{Ensembles of $\mathbb{X}$ with zero mean (Gaussian distributions)\label{zeromeanGX}}

Turning from a specific $\mathbb{X}$ to ensembles of them, these
considerations allow us to predict the prominent differences between the
epdfs in Fig.~\ref{gap-fig2} and Fig.~\ref{gap-fig3}, as well as to understand a number of features in
Fig.~\ref{gap-fig1}. In particular, let us first consider the simplest case: a chiral
ensemble in which $\mathbb{X}$ consists of Gaussian distributed elements with zero mean and finite variance $\sigma ^{2}$. For $%
N\rightarrow \infty $, the epdf obeys the celebrated semicircle law
\cite{ForBook}: $p\left( \eta \right) \propto \sqrt{R^{2}-\eta
^{2}}$ with $R=2\sigma \sqrt{N}$. In Fig.~\ref{gap-fig2}, the red line shows data for the
epdf of the $N=100$ case, with $\sigma =1/2$ and support $\left[ -10,10%
\right] $. Apart from finite $N$ effects (e.g., tails beyond $\pm 10$,
visible peak at $0$), they fit well into the semicircle law (black dashed
line). 
Note that the numerical results for the epdfs in all figures were calculated
from ensembles of $10^7$ random matrices with bin widths of $10^{-3}$.
Also note that, in all figures, lines connect the discrete data points. 
However, we have not drawn lines to connect the data points at $0$ 
and $-1$ with the points inside the gap to emphasize the discontinuity 
in the description of the epdf at the edges of the gap.

Next, we follow the analysis in Section~\ref{specX} and consider the effects of
\textquotedblleft adding $\mathbb{M}$\textquotedblright\ to this ensemble.
In particular, for Fig.~\ref{gap-fig3} shows the epdf of XIE matrices $\left\{ \mathbb{A}%
_{S}\right\} $, using the same zero mean Gaussian ensemble of $\mathbb{X}$
as in Fig.~\ref{gap-fig2}. Thus, we know how the $\rho $'s are distributed: 
\begin{equation*}
p\left( \rho \right) =p\left( \eta \left( \rho \right) \right) \left\vert 
\frac{d\eta }{d\rho }\right\vert \propto \left\vert 1+2\rho \right\vert 
\sqrt{\frac{N}{\rho ^{2}+\rho }-1}
\end{equation*}%
This is the black dashed line plotted in Fig.~\ref{gap-fig3}. The emergence of the gap is
now clear: The solutions to $\rho ^{2}+\rho =\mu =\eta ^{2}\in \left[ 0,N%
\right] $ are less than $\rho _{1-}=-1$ and greater than $\rho _{1+}=0$. The
divergence at the edges of the gap (the Viking-ship like feature) is due to $%
\left( \rho ^{2}+\rho \right) ^{-1/2}$ in the Jacobian. Meanwhile, since the 
$\lambda $'s are squeezed between successive $\rho $'s, the distribution of
the bulk values are identical as $N\rightarrow \infty $, \textit{apart} from
two exceptions to be discussed below. As we see in Fig.~\ref{gap-fig3}, there is excellent
agreement between the data and this prediction: $p_{bulk}\left( \lambda
\right) =p\left( \rho \right) $. Notice that, while $p\left( \eta \right) $
vanishes at $\pm 10$, the bulk of $p\left( \rho \right) $ vanishes at $%
-10.51 $ and $9.51$, the solutions to $\rho ^{2}+\rho =N=100$. These values
provide excellent approximations to the data.

There are two exceptions to $\lambda \simeq \rho $, shown in the insets of
Fig.~\ref{gap-fig3}.
One is the largest eigenvalue, $\lambda _{N+}$, which is not only greater
than $\rho _{N+}=O\left( N^{1/2}\right) $, but \textit{much} greater. If we
assume $\lambda _{N}=O\left( N\right) $, then we can estimate it as follows.
In (\ref{master}), let $1/\left( \lambda _{N}-\rho _{\alpha +}\right) \simeq
1/\lambda _{N}$, so that 
\begin{equation*}
\lambda _{N}\simeq \sum_{\alpha }c_{\alpha }^{2}
\end{equation*}%
But, $c_{\alpha }^{2}\equiv \left\langle u|w_{\alpha }\right\rangle
^{2}=\left\langle u|w_{\alpha }\right\rangle \left\langle w_{\alpha
}|u\right\rangle $ and $\sum_{\alpha }\left\vert w_{\alpha }\right\rangle
\langle w_{\alpha }|=\mathbb{I}$, so that $\sum_{\alpha }c_{\alpha
}^{2}=\left\langle u\right\vert \mathbb{I}\left\vert u\right\rangle =N$.
Thus, to leading order, we find%
\begin{equation*}
\lambda _{N}\simeq N
\end{equation*}%
which justifies our assumption of $\lambda _{N}=O\left( N\right) $. To
account for the next order, we may expand the denominators on the right in (%
\ref{evsEqn}):%
\begin{eqnarray}
\lambda _{N}+1 &=&\sum c_{\alpha }^{2}\left[ 1+\frac{\mu _{\alpha }}{\lambda
_{N}^{2}+\lambda _{N}}+...\right] \\
&=&N+\frac{1}{\lambda _{N}^{2}+\lambda _{N}}\sum \mu _{\alpha }c_{\alpha
}^{2}+...
\end{eqnarray}%
But $\Sigma _{\alpha }\left\vert w_{\alpha }\right\rangle \mu _{\alpha
}\langle w_{\alpha }|=\mathbb{W}$, so that the sum in the last equation is $%
\langle u|\mathbb{X}^{T}\mathbb{X}\left\vert u\right\rangle $. Averaged over
the ensemble, this quantity is well approximated by the sum of the variance
of each element in $\mathbb{X}$. The end result is 
\begin{equation*}
\lambda _{N}=N-1+\sigma ^{2}+...
\end{equation*}%
which is $99.25$ in the case of our simulation study. This approximation
compares well with the distribution of the largest eigenvalue, shown in the
right inset of Fig.~\ref{gap-fig3}.

The other exception is 
$\lambda _{-1}$ 
which lies within the gap: $\left[
\rho _{1-},\rho _{1+}\right] =\left[ -1,0\right] $. While a solution to (\ref%
{master}) must exist between these $\rho $ values, finding an analytic
expression of its precise location remains a challenge. As the inset on the
left in Fig.~\ref{gap-fig3} shows, it appears to be \textquotedblleft tightly
bound\textquotedblright\ to the \textit{right} edge with probability
decreasing exponentially as $\lambda $ decreases from $0$. Such behavior is
opposite to the case in Fig.~\ref{gap-fig1} (\textquotedblleft tight
binding\textquotedblright\ to the \textit{left} edge). Below, we will return to
further investigations of this eigenvalue in the gap.

\begin{figure}
\centering
\includegraphics[width=\textwidth]{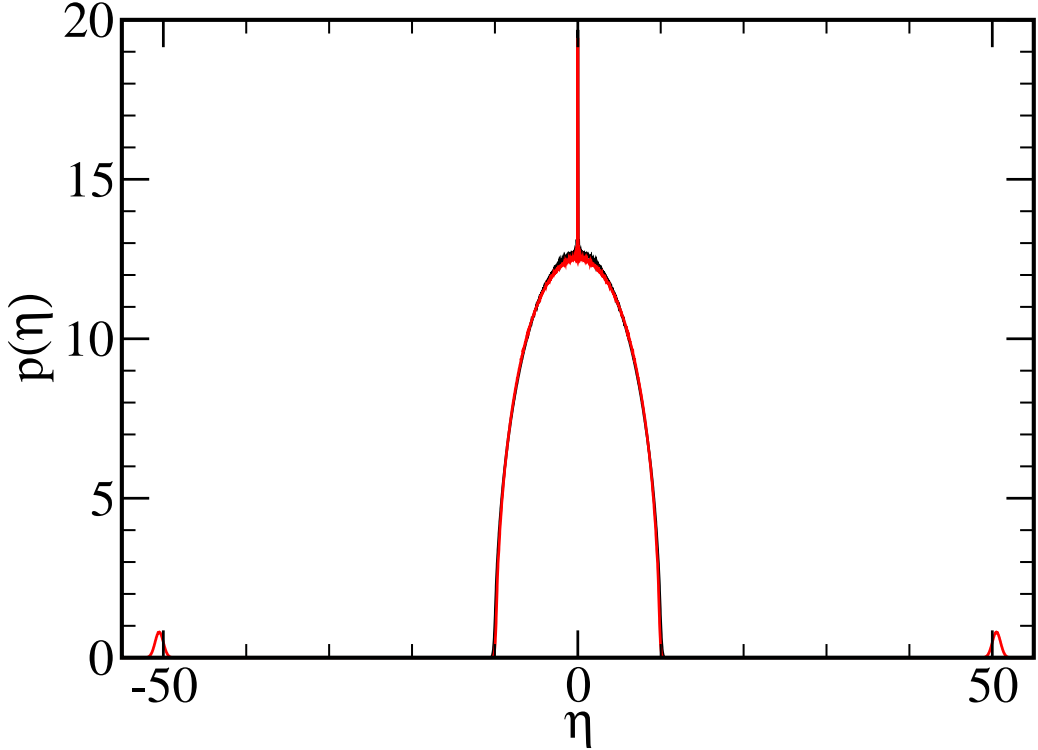}
\caption{Eigenvalue probability distribution function of $2N \times 2N$ chiral matrices with $N=100$
and Gaussian distributed elements with variance of 0.25 and mean of zero (black) and of 0.5 (red).  
Note the two eigenvalues that separate from the bulk when the mean is non-zero.
}
\label{gap-fig4}
\end{figure}

\subsection{Ensembles of $\mathbb{X}$ with positive mean (Boolean
distributions)}

A well-known phenomenon in random matrix theory is that, if it consists of
i.i.d. random elements with non-zero mean, an isolated eigenvalue may emerge
from the bulk (e.g, the semi-circle) epdf. This is certainly the case if the
elements were Boolean variables, i.e., randomly $1$ with probability, $q$,
and $0$ with probability $1-q$. The mean is $q$ and the variance is $\sigma
^{2}=q\left( 1-q\right) $. By choosing $q=1/2$ in simulations, universality
guarantees that the bulk part of the epdf should, as $N\rightarrow \infty $,
approach that in the Gaussian example above. The only difference is the
presence of isolated eigenvalues, separated from the bulk~\cite{BFF2009},
located at $\pm \eta _{N}$ ($\simeq 50.5$ here). For the convenience of the
readers, we reproduce this behavior in Fig.~\ref{gap-fig4}, so that it can be compared to
Fig.~\ref{gap-fig1}. Furthermore, such a large $\eta _{N}$ is connected with the (lower
doublet of) its associated eigenvector, $\left\vert w_{N}\right\rangle $,
being mainly along $\left\vert u\right\rangle $.

To appreciate these connections, consider the simplest case, namely, adding
mean $q\rightarrow \infty $ to $\mathbb{X}_{G}$ (the notation we use here
for a zero mean Gaussian ensemble such as in Section~\ref{zeromeanGX}, with
variance $\sigma ^{2}=O\left( 1\right) $). Denoting $\mathbb{X}$ by $\mathbb{%
X}_{G}+q\left\vert u\rangle \langle u\right\vert $, we examine (\ref{W ev}). 
\begin{equation}
\mathbb{X}^{T}\mathbb{X}\left\vert w_{N}\right\rangle =\left( \mathbb{X}%
_{G}^{T}+q\left\vert u\rangle \langle u\right\vert \right) \left( \mathbb{X}%
_{G}+q\left\vert u\rangle \langle u\right\vert \right) \left\vert
w_{N}\right\rangle =\mu _{N}\left\vert w_{N}\right\rangle 
\end{equation}%
Clearly, to leading order, $\left\vert w_{N}\right\rangle \propto \left\vert
u\right\rangle $ and $\mu _{N}=\left( qN\right) ^{2}$. Normalization leads
to $\left\vert w_{N}\right\rangle =\left\vert u\right\rangle /\sqrt{N}$,
along with $c_{N}=\sqrt{N}$. We can find the next order correction of the
eigenvalue by standard means, regarding $\mathbb{V}\equiv q\left\vert
u\rangle \langle u\right\vert \mathbb{X}_{G}+q\mathbb{X}_{G}^{T}\left\vert
u\rangle \langle u\right\vert +\mathbb{X}_{G}^{T}\mathbb{X}_{G}$ as a
perturbation for $\mu _{N}$. To first order, we find $\mu _{N}=\left(
qN\right) ^{2}+\langle w_{N}|\mathbb{V}\left\vert w_{N}\right\rangle
+...=\left( qN\right) ^{2}+\sigma ^{2}N+...$. Notice that the $O\left(
q\right) $ terms average to zero and so, the next non-vanishing contribution
is $O\left( 1\right) $. Thus, if that is to be kept, we should consider
second order perturbation as well. At that order $\mathbb{V}^{2}$ contains
terms $\propto $ $\left( qN\right) ^{2}$ which do not average to zero. A
straightforward but tedious computation leads to another $\sigma ^{2}N+...$
The final result is $\mu _{N}=\left( qN\right) ^{2}+2\sigma ^{2}N+...$, so
that $\eta _{N+}=\sqrt{\mu _{N}}=qN+\sigma ^{2}/q+...$. Remarkably, it is in
excellent agreement with preliminary simulations studies using $N=100$, $%
\sigma =1/2,1,2$, and $q=1/2,1,2$. Of course, the typical limit of interest
is large $N$ with fixed $q$, which does not necessarily commute with the
large $q$ fixed $N$ limit here. Nevertheless, this result is the same as
implicit predictions in the literature (See, e.g., \cite{BFF2009}).

Meanwhile, since $\Sigma _{\alpha }c_{\alpha }^{2}=N$,
we conclude that the sum over the rest of the $\alpha $'s must be 
\begin{equation*}
\Sigma ^{\prime }\equiv \sum_{\alpha =1}^{N-1}c_{\alpha }^{2}=O\left(
1\right)
\end{equation*}%
Thus, on the average, each $c_{\alpha }^{2}$ is expected to be $O\left(
1/N\right) $. As $c_{\alpha }=O\left( N^{-1/2}\right) $ is the projection of 
$\left\vert w_{\alpha }\right\rangle $ onto $\left\vert u\right\rangle $, we
conclude the obvious, i.e., the rest of the eigenvectors lie mainly in a
subspace orthogonal to $\left\vert u\right\rangle $.

Finally, we turn to the understanding of the epdf in Fig.~\ref{gap-fig1}. One possible
route is to consider the effects of adding\ a mean to the $\mathbb{X}_{G}$'s
in (the off diagonal blocks of) into the $\mathbb{A}_{S}$ associated with
Fig.~\ref{gap-fig3}. Here, we purse the easier route, examine the effects of
\textquotedblleft adding $\mathbb{M}$\textquotedblright\ to the $\mathbb{A}%
_{B}$'s associated with Fig.~\ref{gap-fig4} (i.e., from adjacency matrices of bipartite
graphs with Boolean $\mathbb{X}$'s to the same for split graphs). We will
find that most of striking features in Fig.~\ref{gap-fig1} can be understood along the
these lines as in Sec.~\ref{zeromeanGX} above.

First, as expected, the dominant part of the epdf -- the bulk -- are
affected in much the same way that Fig.~\ref{gap-fig2} is transformed into Fig.~\ref{gap-fig3}. Thus,
we see that this part of $p\left( \lambda \right) $ vanishes outside $\simeq
-10.51$ and $\simeq 9.51$, as well the presence of a gap between $\simeq -1$
and $0$. Turning to the separated eigenvalues outside, note that their
distributions peak near $-22.10$ and $120.06$. Both are quite far from the $%
\pm 50.5$ values in Fig.~\ref{gap-fig4}. To provide a good estimate for these, we turn
again to (\ref{master}), but must account for the separated eigenvalues in
Fig.~\ref{gap-fig3}
($\eta _{N\pm }$) being $O\left( N\right) $,
instead of $\eta _{N\pm }=O\left( N^{1/2}\right) $ in Fig. 2. Starting with
Eqn (\ref{evsEqn} and seeking a $\lambda _{N}=O\left( N\right) \gg \eta
_{\alpha <N}$, we find%
\begin{eqnarray*}
1 &=&\lambda _{N}\sum \frac{c_{\alpha }^{2}}{\lambda _{N}^{2}+\lambda
_{N}-\mu _{N}}=\lambda \left\{ \frac{c_{N}^{2}}{\lambda _{N}^{2}+\lambda
_{N}-\mu _{N}}+\sum_{\alpha =1}^{N-1}\frac{c_{\alpha }^{2}}{\lambda
_{N}^{2}+\lambda _{N}-\mu _{\alpha }}\right\}  \\
&=&\frac{\lambda _{N}\left( N-\Sigma ^{\prime }\right) }{\lambda
_{N}^{2}+\lambda _{N}-\mu _{N}}+\frac{\Sigma ^{\prime }}{1+\lambda _{N}}=%
\frac{\lambda _{N}N}{\lambda _{N}^{2}-\mu _{N}}+O\left( \frac{1}{N}\right) 
\end{eqnarray*}%
Substituting the leading order approximation for $\mu _{N}$, and solving 
$\lambda _{N}^{2}-\lambda _{N}N-q^{2}N^{2}=0$, we arrive at%
\begin{equation*}
\lambda _{N\pm }=\frac{N}{2}\left[ 1\pm \sqrt{1+4q^{2}}\right] 
\end{equation*}%
which are $N\left( 1\pm \sqrt{2}\right) /2\simeq -20.7,120.7$ for the
Boolean case. These are remarkably close to the observed values, given that
we took into account only the leading order!

Second, there should be an eigenvalue between $\eta _{\left( N-1\right) +}$
and $\eta _{N+}$, but the data implies it is \textquotedblleft tightly
bound\textquotedblright\ to the bulk. This aspect can be understood by the
following illustration. Consider a secular equation with just two terms, 
$1=r_{1}/\left( \lambda -\rho _{1}\right) +r_{2}/\left( \lambda -\rho
_{2}\right) $, and let $r_{1}=O\left( 1/\sqrt{N}\right) $, $\rho
_{1}=O\left( \sqrt{N}\right) $, while $r_{2},\rho _{2}=O\left( N\right) $. A
simple plot of the right will provide an intuitive picture for the behavior
of solutions. A recursive form 
\begin{eqnarray*}
\lambda _{i}&=&\rho _{i}+r_{i}+r_{j\neq i}\left( \lambda -\rho _{i}\right)
/\left( \lambda -\rho _{j\neq i}\right) \\
&\simeq& \rho _{i}+r_{i}+r_{j\neq
i}\left( r_{i}+...\right) /\left( \rho _{i}+r_{i}-\rho _{j\neq i}+...\right)
\end{eqnarray*}%
provides a more quantitative estimate, as we see the effect of our
assumptions, namely, $\lambda _{1}=\rho _{1}+O\left( 1/\sqrt{N}\right) $
being \textquotedblleft tightly bound\textquotedblright\ to $\rho _{1}$,
while $\lambda _{2}=\rho _{2}+r_{2}+...$) can be far from $\rho _{2}$.

\begin{figure}
\centering
\includegraphics[width=\textwidth]{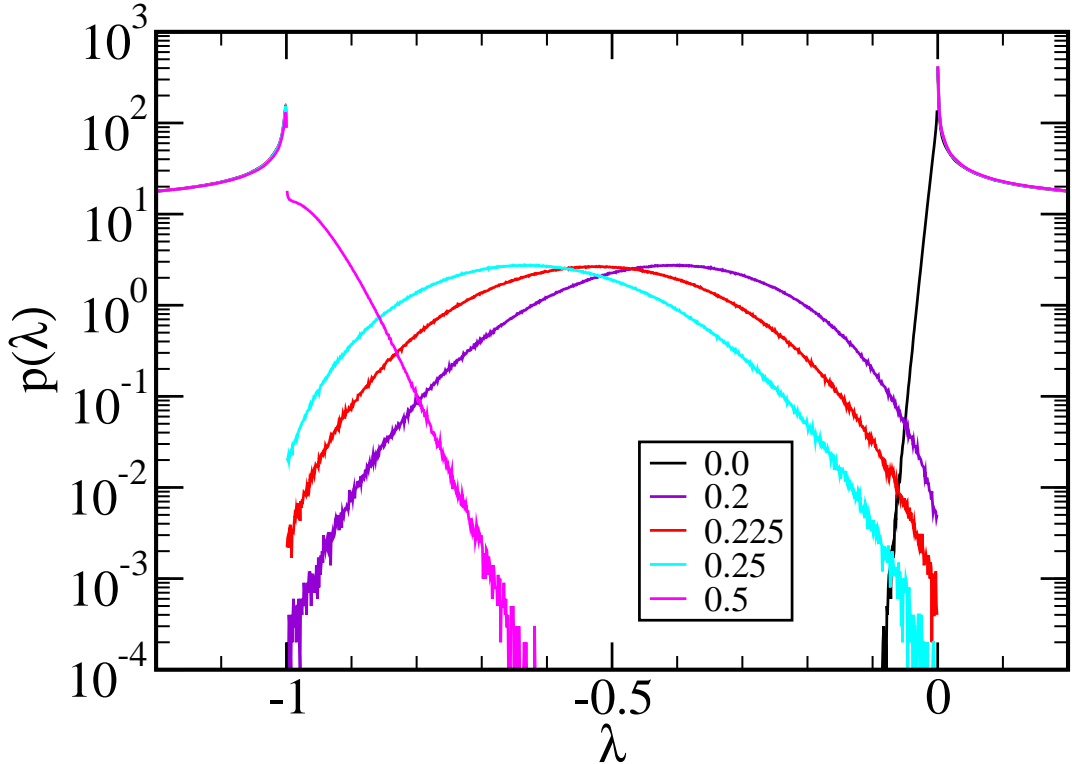}
\caption{The effect of the mean value (of the i.i.d. random matrox elements) on the probability distribution function of the gap-eigenvalue, $\lambda_{1-} \in \left[ -1,0\right]$. (Legend lists the mean values.) These studies are based on 200x200 XIE matrices with Gaussian distributed elements of variance of 0.25.}
\label{gap-fig5}
\end{figure}

\begin{figure}
\centering
\includegraphics[width=\textwidth]{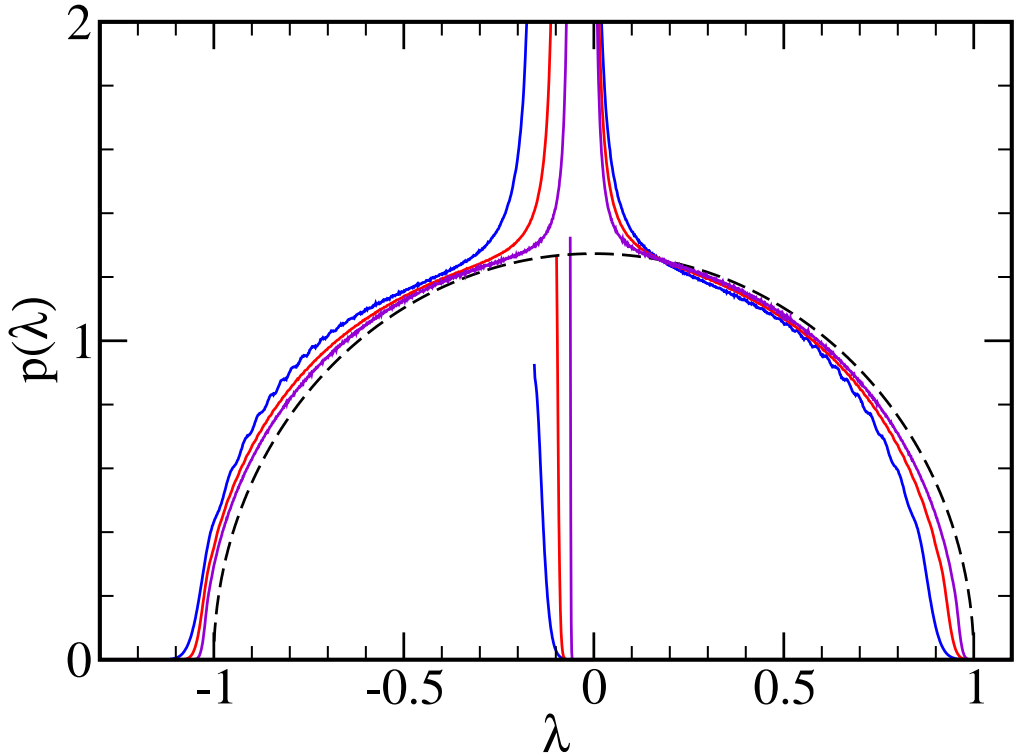}
\caption{Scaled eigenvalue probability distribution functions of $2N \times 2N$ 
random split graph adjacency matrices with I-E connection probability 0.5 and $N=40$ 
(blue), $N=100$ (red) and $N=250$ (violet). 
Black dashed line shows limiting semi-circle distribution.
Note the almost vertical line at the left edge of each of the gaps indicate that there is one eigenvalue
\textquotedblleft tightly bound\textquotedblright\ to the left edge in these cases.
}
\label{gap-fig6}
\end{figure}

What remains is perhaps the most challenging task: predicting the
location and distribution of the eigenvalue in the gap, $\lambda _{1-}$. As
noted in the illustration above, the location of the solutions depend on the
details of the parameters. In our case, there are many poles nearby and a
range of residues, so that $\lambda _{1-}$ can be anywhere between $\rho
_{1-}\rightarrow -1$ and $\rho _{1+}\rightarrow 0$. So far, we have been
able to explore this issue only through simulations. As Fig.~\ref{gap-fig5} shows, it can
be tightly bound to either edge of the gap, or be isolated from both. In
this case, we varied only the mean (of Gaussian $\mathbb{X}$ with variance $%
1/4$), from $0$ to $0.5$ (corresponding to the cases of Fig.~\ref{gap-fig3} and \ref{gap-fig1},
respectively). We see that $\lambda _{1-}$ moves from being bound to $0$,
through being detached at intermediate values and to being bound near $-1$.
Indeed, the distribution near the right edge appears to be a pure
exponential. On closer examination, the weight of this component appears to be less
than unity, indicating that $\lambda _{1-}$ is found to be positive for some
realizations of $\mathbb{X}$. It is unclear if this feature persists as $
N\rightarrow \infty $; further studies are underway to explore if the
exponential distribution here approaches a universal limit. In the language
of interactions between eigenvalues, the
interpretation would be that $\lambda _{1-}$ experiences a constance
attractive force from a wall (formed by the the bulk eigenvalues). However,
the potential associated with such a wall is finite, as $\lambda _{1-}$ is
able to penetrate into the wall with non-zero probability. As the mean
increases, the distribution of the detached $\lambda _{1-}$ appears to be
more Gaussian-like, with noticeable asymmetry. A possible conjecture is
that, as $N\rightarrow \infty $, the width of this Gaussian decreases,
resulting in a $\delta $-like distribution. In the last case with mean $0.5$%
, the distribution is reminiscent of a critical point associated with an
unbinding transition: with $\ln p$ being linear far from the wall, but
quadratic near $-1$. Clearly, many questions arise and will provide fertile
grounds for future research.

\section{Summary and outlook}

With application in a wide range of fields, ensembles of random matrices and
the pdf of their associated spectra\ are fascinating topics of study. For
example, in network science and graph theory, adjacency matrices of random
networks/graphs form such ensembles. In this paper, we focus on the
connection between ensembles of bipartite graphs and the related split
graphs. Our motivation comes from social networks, where bipartite graphs
describe the contacts between two otherwise totally disjoint subgroups,
while split graphs correspond to connecting all members of one of the two
subgroups (to form a \textquotedblleft clique\textquotedblright ).
Simulations show several striking features in the epdf which result from
adding such contacts, especially the emergence of a gap between $-1$ and $0$
in the spectra.%
\footnote{%
Though our study was based on the symmetric case where the numbers in the two 
subgroups are the same ($N_1=N_2$), the generalization to asymmetric cases 
is straightforward, with similar conclusions. The only difference is that all 
the random matrices will display a null space of dimension $| N_1 - N_2 |$.}
Our conclusion is that, though nearly all of these features
can be understood, many lines of inquiry remain open. We end by listing a
few here.

Much of our analysis can be improved, both in terms of rigor and accuracy
beyond the leading order. The most pressing issue is a better understanding of
the location and distribution of the gap eigenvalue. The language of
interacting eigenvalues should be explored in this context, so as to clarify
if there are phase transitions like unbinding and if there are anomalous
properties. It is known that certain behavior is universal in the large $N$
limit. 
Some preliminary data, presented in Fig.~\ref{gap-fig6}, show the convergence of the \textit{%
bulk} distribution, as $N$ increases, towards a semi-circle. 
What are the finite size effects? and is there finite size scaling?
Work in this direction is in progress and the
results will be interesting regardless of what they show. By contrast, we
know only of the presence of the gap eigenvalue, while much of the details
of the location and distribution remain to be explored. Beyond the issues
associated with i.i.d. elements in our $\mathbb{X}$'s, a much more
challenging problem is the effects of correlations. In our original model of
extreme introverts and extroverts, the dynamic rules are simple: choose an
individual at random and let it cut/add a random link. Yet, the resulting
ensemble of adjacency matrices, $\mathcal{P}\left( \mathbb{A}_{XIE}\right) $%
, contains highly correlated elements\cite{LSZ2014}. Indeed, as the numbers
of one subgroup overtake those of the other, there is an extraordinary
transition \cite{ZLS2012,BLSZ2015}. At the transition (when the numbers are
the same), there are giant fluctuations in $\mathbb{X}$, specifically,
$\langle u| \mathbb{X}\left\vert u\right\rangle =\sum_{ij}X_{ij}$ being
equally likely to be in the entire range $\left( 0,N^{2}\right) $. 
How are such giant 
fluctuations and correlations reflected in the epdfs of the adjacency
matrices? Preliminary data show qualitatively the same features as in Fig.~\ref{gap-fig1}~\cite{Ezzat}, 
but a systematic study may find novel and interesting behavior.
Further, while the studies here involve explicitly \textit{known} ensembles,
the more realistic models of social networks involve introverts and
extroverts who \textquotedblleft prefer\textquotedblright\ generic, finite
degrees. In that case, the dynamics typically violates detailed balance, so
that the stationary distributions of the adjacency matrices are \textit{not}
known analytically\cite{LJSZ2013}. What features are displayed in those
epdfs? and can we understand them, whether they belong to the same
universality classes or not? Such questions take us to the vast and
unfamiliar territory of non-equilibrium statistical mechanics, with
seemingly unlimited and novel phenomena to be discovered.

\section*{Acknowledgements}

One of us (RKPZ) has benefited much from numerous discussions with John
Cardy, over many years, on various topics in theoretical physics. It is a
great pleasure for us to contribute this article to a special issue of this
journal in honor of his 70th birthday. This research is support in part by
the United States NSF through grant DMR-1507371. 
This study germinated during the authors' stay at the MPIPKS
in 2013-2014, under the auspices of an ASG. The hospitality of Frank
J\"{u}licher is gratefully acknowledged. Finally, we thank Uwe T\"{a}uber
for inviting us to write an article for this occasion, as well as 
Zoltán Toroczkai and Mohammadmehdi Ezzatabadipour for illuminating discussions.

\section{Appendix}

Here, we consider the exceptions to the generic case detailed in Section 2.

\subsection{Orthogonality}

If some of the eigenvector of $\mathbb{W}$ are orthogonal to $\left\vert
u\right\rangle $ (denoted by $|w_{o}\rangle $), then $c_{o}=0$ and Eqn. (\ref%
{v}) implies that $\left\vert v\right\rangle $ is just $|w_{o}\rangle $ with 
$\lambda _{o\pm }=\rho _{o\pm }$. Clearly, these are the cases where $c=0$.
The associated eigenvectors for $\mathbb{A}_{S}$ are the doublets $\mathbb{W}%
|w_{o}\rangle /\lambda _{o\pm },|w_{o}\rangle $. This leaves an orthogonal
subspace spanned by $c\neq 0$ vectors, which we label by $\left\vert
w_{\gamma }\right\rangle $; $\gamma =1,...,L<N$. Following the same route as
above, we have 
\begin{equation}
\mathbb{W}\left\vert w_{\gamma }\right\rangle =\mu _{\gamma }\left\vert
w_{\gamma }\right\rangle ;~~c_{\gamma }=\left\langle u|w_{\gamma
}\right\rangle \neq 0
\end{equation}%
and we can decompose 
\begin{equation}
\mathbb{W}=\mathbb{W}_{o}+\mathbb{W}_{\gamma }\equiv \sum_{o}\mu
_{o}\left\vert w_{o}\rangle \langle w_{o}\right\vert +\sum_{\gamma }\mu
_{\gamma }\left\vert w_{\gamma }\rangle \langle w_{\gamma }\right\vert 
\end{equation}%
Within this subspace, we again have%
\begin{equation}
\mathbb{W}_{\gamma }\left\vert v\right\rangle =\left( \lambda ^{2}+\lambda
\right) \left\vert v\right\rangle -\lambda c\left\vert u\right\rangle 
\end{equation}%
The rest of the analysis is now clear and and (\ref{evsEqn}) now reads%
\begin{equation}
1=\lambda \sum_{\gamma =1}^{M}\frac{c_{\gamma }^{2}}{\lambda ^{2}+\lambda
-\mu _{\gamma }}
\end{equation}%
leading us to the rest of the $2L$ eigenvalues and eigenvectors of $\mathbb{A%
}_{S}$.

In the special case that all but one $|w_{\alpha }\rangle $ is orthogonal to 
$\left\vert u\right\rangle $, then $\left\vert u\right\rangle $ itself must
also be an eigenvector. In that case, let us denote $\mathbb{W}\left\vert
u\right\rangle =\mu _{u}|u\rangle $ which leads us to 
\begin{equation}
\mathbb{W}\left\vert u\right\rangle =\left( \lambda _{u}^{2}+\lambda
_{u}-\lambda _{u}N\right) \left\vert u\right\rangle 
\end{equation}%
and%
\begin{equation}
\lambda _{u\pm }=\frac{1}{2}\left\{ N-1\pm \sqrt{\left( N-1\right) ^{2}+4\mu
_{u}}\right\} 
\end{equation}%
The reduction to $N=1$ is trivial, as $\mathbb{A}_{B}\equiv \mathbb{A}_{S}$
and $\lambda _{\pm }\equiv \eta _{\pm }$! Also, it means that the average of
the elements of $\mathbb{W}$ is just $\mu _{u}/N$, since that average is $%
\sum_{ij}W_{ij}/N^{2}=\langle u|\mathbb{W}\left\vert u\right\rangle /N^{2}$.

\subsection{Null space}

Suppose $K$ of the $\mu $'s are zero (corresponding to $2L$ $\eta $'s),
denoted by $\mu _{\kappa }=0;\kappa =1,...,K<N$. and associated with
eigenvectors $\mathbb{W}\left\vert w_{\kappa }\right\rangle =0$. Since $%
\mathbb{W=X}^{T}\mathbb{X}$, we must have $\mathbb{X}\left\vert w_{\kappa
}\right\rangle =0=\mathbb{X}^{T}\left\vert \tilde{w}_{\kappa }\right\rangle $%
, where$\left\vert \tilde{w}_{\kappa }\right\rangle $ is the transpose of
the left eigenvector of $\mathbb{X}$. In other words, the doublets for $%
\mathbb{A}_{B}$ associated with $\eta _{\kappa \pm }=0$ are $\left(
\left\vert 0\right\rangle ,\left\vert w_{\kappa }\right\rangle \right) $ and 
$\left( \left\vert \tilde{w}_{\kappa }\right\rangle ,\left\vert
0\right\rangle \right) $. Now, these $\eta $'s lead us to $\rho _{\kappa +}=0
$ and $\rho _{\kappa -}=-1$. The first naively imply the pole terms there
are absent (as its residue is $\rho _{\kappa +}$). However, its treatment is
similar to the orthogonal cases. In particular, we simply verify that the
doublets $\left( \left\vert \tilde{w}_{\kappa }\right\rangle ,\left\vert
0\right\rangle \right) $ is in the null space of $\mathbb{A}_{S}$, i.e.,
they are the eigenvectors of $\mathbb{A}_{S}$ associated with $\lambda
_{\kappa +}=0$. As for $\rho _{\kappa -}=-1$, the pole terms are
non-singular and the treatment for them remains the same as above. This
subsection also addresses the issue when $\lambda $ vanishes. 

\subsection{Degeneracy}

If there is a set of eigenvectors, $\left\vert w_{s}\right\rangle $, with
the same eigenvalue $\mu _{s}$, then the only point is that the residue of
the poles at $\rho _{s\pm }$ becomes $\Sigma _{s}c_{s}^{2}=\langle u|\left[
\Sigma _{s}\left\vert w_{s}\right\rangle \langle w_{s}|\right] \left\vert
u\right\rangle =\langle u|\mathbb{I}_{s}\left\vert u\right\rangle $, where $%
\mathbb{I}_{s}$ is the unit matrix within the subspace spanned by $\left\{
\left\vert w_{s}\right\rangle \right\} $. Thus, it is invariant to rotations
within that subspace, i.e., $\Sigma _{s}c_{s}^{2}$ does not depend on the
precise choice of the set $\left\{ \left\vert w_{s}\right\rangle \right\} $.


\bigskip

\section*{References}
\begin{thebibliography}{10}
\bibitem{PorterBook} C. E. Porter, {\it Statistical theories of spectra: Fluctuations}, (Academic, New York, 1965).
\bibitem{W1928} J. Wishart, Biometrika {\bf 20A}, 32 (1928).
\bibitem{May1972} R. M. May, Nature (London) {\bf 238} 413 (1972).
\bibitem{May1973} R. M. May, {\it Stability and Complexity in Model Ecosystems} (Princeton University Press, Princeton, 1973).
\bibitem{FSB2003} P. J. Forrester, N. C. Snaith, and J. J. M. Verbaarschot, J. Phys. A: Math. Gen. {\bf 36}, R1 (2003).
\bibitem{ForBook} P. F. Forrester, {\it Log-Gases and Random Matrices}, (Princeton University Press, Princeton, 2010).

\bibitem{Wigner1955} E. Wigner, Ann. of Math. {\bf 62}, 548 (1955).
\bibitem{Wigner1958} E. Wigner, Ann. of Math. {\bf 67}, 325 (1958).
\bibitem{a1971} L. Arnold, Z. Wahrscheinlichkeitstheorie und Verw. Gebiete {\bf 19}, 191 (1971).
\bibitem{MP1967} V. A. Marchenko, and L. A. Pastur, Mat. Sb. N.S. (in Russian) {\bf 72}, 507 (1967).


\bibitem{BGN2011} F. Benaych-Georges and R. R. Nadakuditi, Adv. Math. {\bf 227}, 494 (2011).
\bibitem{OW2017} S. O'Rourke and P. M. Wood, Ann. Inst. H. Poincar\'{e} Probab. Statist. {\bf 53}, 1241 (2017).

\bibitem{cheung1994} F. Cheung, {\it Spectral Graph Theory}, (no. 92 in CBMS Regional Conference Series. Conference Board of Mathematical Sciences, 1994).
\bibitem{BJ2007} J.N. Bandyopadhyay and S. Jalan, Phys. Rev. E {\bf 76}, 026109 (2007).
\bibitem{mieghem2011} P. Mieghem, {\it Graph Spectra for Complex Networks}, (Cambridge University Press, New York, 2011). 

\bibitem{nn2012} R. R. Nadakuditi and M. E. J. Newman,  Phys. Rev. Lett.  {\bf 108}, 188701  (2012).
\bibitem{p2013} T. P.  Peixoto, Phys. Rev. Lett.  {\bf 111}, 098701  (2013).

\bibitem{MSA2008} B.D. MacArthur, R.J. Sánchez-García, and R.W. Anderson, Discrete Appl. Math. {\bf 156}, 3525 (2008). 
\bibitem{NGB2015} A. Nyberg, T. Gross, and K. E. Bassler, J. Comp. Networks {\bf 3}, 543 (2015).
\bibitem{DK2017} C.P. Dettmann and G. Knight, preprint arXiv:1704.00640 (2017).

\bibitem{ADKYZ2008} A. Arenas, A. Díaz-Guilera, J. Kurths, Y. Moreno, and C. Zhou, Phys. Rep. {\bf 469}, 93 (2008).
\bibitem{MS2009} B.D. MacArthur, and R.J. Sánchez-García, Phys. Rev. E {\bf 80}, 026117 (2009). 

\bibitem{clv2003} F. Chung, L. Lu, and V. Vu, Proc. Natl. Acad. Sci. U.S.A. {\bf 100}, 6313 (2003).

\bibitem{L1964} D.W. Lang, Phys. Rev. {\bf 135}, B1082 (1964).
\bibitem{JKT1968} R.C. Jones, J.M. Kosterlitz, and D.J. Thouless, J. Phys. A {\bf 11}, L45 (1978).

\bibitem{BFF2009} K. E. Bassler, P. J. Forrester, and N. E. Frankel, J. Math. Phys. {\bf 50}, 033302 (2009).

\bibitem{er1959} P. Erdős and A. Rényi, Publicationes Mathhematice (Debrecen) {\bf 6},  290 (1959).

\bibitem{FK1981} Z. Füredi and J. Komlós, Combinatorica {\bf 1}, 233 (1981).

\bibitem{RCKT2008} T. Rogers, I. P. Castillo, R. Kühn, and K. Taked, Phys. Rev. E {\bf 78}, 031116 (2008).

\bibitem{fdbv2001} I. J. Farkas, I. Derényi, A.-L. Barabási, and T. Vicsek, Phys. Rev. E {\bf 64}, 026704 (2001).

\bibitem{ZLJS2011} R. K. P. Zia, W. Liu, S. Jolad, and B. Schmittmann, Phys. Proc. {\bf 15}, 102 (2011).
\bibitem{ZLS2012} R. K. P. Zia, W. Liu, and B. Schmittmann, Phys. Proc. {\bf 34} 124 (2012).
\bibitem{LSZ2012} W. Liu, B. Schmittmann, and R. K. P. Zia, Europhys. Lett. {\bf 100}, 66007 (2012).
\bibitem{LJSZ2013} W. Liu, S. Jolad, B. Schmittmann, and R. K. P. Zia, J. Stat. Mech. P08001 (2013).
\bibitem{LSZ2014} W. Liu, B. Schmittmann, and R. K. P. Zia, J. Stat. Mech. P05021 (2014).

\bibitem{FH1977} S. Földes, and P. L. Hammer, Canad. J. Math. {\bf 29}, 666 (1977).
\bibitem{TC1979} R. Tyshkevich, and A. A. Chernyak, Vesti Akad. Navuk BSSR Ser. Fiz.-Mat. Navuk {\bf 5}, 14 (1979).

\bibitem{BLSZ2015} K. E. Bassler, W. Liu, B. Schmittmann, and R. K. P. Zia, Phys. Rev. E {\bf 91}, 042102 (2015).
\bibitem{BDZ2015} K. E. Bassler, D. Dhar, and R. K. P. Zia, J. Stat. Mech. P07013 (2015).

\bibitem{HLH1983} P.W. Holland, K.B. Laskey, and S. Leinhardt, Soc. Networks {\bf 5}, 109 (1983).
\bibitem{FMW1985} S.E. Fienberg, M.M. Meyer, and S.S. Wasserman, J. Am. Stat. Assoc. {\bf 80}, 51 (1985).
\bibitem{FW1992} K. Faust and S.S. Wasserman, Soc. Networks {\bf 14}, 5 (1992).

\bibitem{BFF2010} K. E. Bassler, P. J. Forrester, and N. E. Frankel, J. Math. Phys. {\bf 51}, 123305 (2010).

\bibitem{Ezzat} M. Ezzatabadipour, K.E. Bassler, and R.K.P. Zia, unpublished. 


\end{thebibliography}
\end{document}